# Energy Dependence of Particle Ratios in High Energy Nucleus-Nucleus Collisions: A USTFM Approach

**Inam-ul Bashir and Saeed Uddin***

*Department of Physics, Jamia Millia Islamia (Central University) New-Delhi, India*

## Abstract

We study the identified particle ratios produced at mid-rapidity in heavy ion collisions, along with their correlations with the collision energy. We employ our earlier proposed Unified Statistical Thermal Freeze-out Model (USTFM), which incorporates the effects of both longitudinal as well as transverse hydrodynamic flow in the hot hadronic system. A fair agreement seen between the experimental data and our model results confirms that the particle production in these collisions is of statistical nature. The variation of the chemical freeze-out temperature and the baryon chemical potential with respect to collision energies is studied. The chemical freeze-out temperature is found to be almost constant beyond the RHIC energy and is found to be close to the QCD predicted phase transition temperature suggesting that the chemical freeze-out occurs soon after the hadronization takes place. The vanishing value of chemical potential at LHC indicates very high degree of nuclear transparency in the collision.

**saeed_jmi@yahoo.co.in*

==================================================================

## Introduction

Relative hadron yields and their correlations are observable which can provide information on the nature, composition and size of the medium from which they originate in high energy heavy ion collisions where a strongly interacting nuclear matter at high energy density and temperatures is formed. Within the framework of the statistical model, it is assumed that a hot and dense fireball is formed over an extended region for a brief period of time ( ~ a few fm/c) after the initial collision which undergoes collective expansion leading to a decrease in its temperature and finally to the hadronization. After the hot fireball formed in such collisions hadronizes, which initially has a very high density of partons (i.e., quarks and gluons), the hadrons keep interacting with each other and the particle number changing (inelastic) reaction processes continue to take place till the temperature (and hence the mean thermal energy) drops to a certain value where a given reaction process almost comes to a stop. Those particle number changing reaction processes ((e.g. strangeness exchange process, etc.) stop earlier for which the threshold energy is larger. The temperature at which the particle number changing process for a given hadron almost stops is called the "chemical freeze-out" temperature of that hadronic specie. However, the (elastic) rescattering still continues to take place and causes a build up the collective (hydrodynamic) expansion. Consequently, the matter becomes dilute and the mean free path for the elastic reaction processes of given hadronic specie becomes comparable with the system size. At this stage the scattering processes stop and the given hadron decouples from the rest of the system. This is called the "kinetic or thermal freeze-out" after which the hadron's energy/momentum spectrum is frozen in time [1]. As the inelastic cross sections are only a small fraction of the total cross section at lower (thermal) energies hence the inelastic processes stop well before the *elastic* ones. Thus chemical freeze-out precedes kinetic or thermal freeze-out [2].

Statistical thermal models have successfully reproduced the essential features of particle production in heavy-ion collisions [3] as well as in many types of elementary collisions [4]. Systematic studies of particle yields using experimental results at different beam energies have revealed a clear underlying freeze-out pattern for particle yields in heavy-ion collisions [5]. The success of the statistical (thermal) models in describing the ratios of hadron yields produced in heavy ion collisions is remarkable. The agreement of the particle ratios with simple predictions of the statistical models is a key argument for the thermalization of the system formed in heavy ion collisions. Measurements of anti-particle to particle ratios in these collisions give information on the net baryon density or baryon chemical potential achieved and are thus of interest in characterizing the environment created in these collisions. It has also been suggested that the measurement of strange anti-baryon to baryon ratios could help distinguish between a hadron gas and deconfined plasma of quarks and gluons [6]. For a boost invariant system at mid-rapidity for the RHIC and LHC energies, the particle yields dN/dy change only by a few percentages in the rapidity window $|y|<1$. The particles ratios detected at mid-rapidity are the integrated yield from various parts of the fireball.

In this paper, we attempt to reproduce the particle ratios and to study their correlations and the energy dependence in the HG scenario by using our phenomenological boost invariant Unified Statistical Thermal Freeze-out Model (USTFM) [1,7] which assumes that at freeze-out all the hadrons in the hadron gas resulting from a high energy nuclear collision follow an equilibrium distribution. The local particle phase space densities have the form of the Fermi-Dirac or Bose-Einstein statistical distributions.

## Results and Discussion

The nuclear matter created in high energy heavy ion collisions is assumed to form an ideal gas that can be described by Grand Canonical Ensemble. The density of the particle $i$ can then be written as,

$$n_i = \frac{g_i}{2\pi^2} \int_0^\infty \frac{k^2 dk}{\exp\left[\frac{E_i(k)-\mu_i}{T}\right] \pm 1} \quad (1)$$

Where $E_i = \sqrt{(k^2 + m_i^2)}$ is the energy, $k$ is the momentum of the particle specie, $g_i = (2J_i + 1)$ is the spin degeneracy factor, $\mu_i$ is the chemical potential of the particle species $i$ and T is the temperature. The (+) sign is for fermions and (-) sign is for bosons. For high temperatures and energies, the Bose or Fermi Dirac statistics can be replaced by the Boltzmann statistics by dropping the ± 1 term. The chemical freeze-out relates with the equilibrium between different flavors. If the hadron gas reaches chemical equilibrium, the particle abundance is described by chemical potentials and temperatures. The information of the chemical freeze-out can be extracted from particle ratios in the measurement. Relative particle production can be studied by particle ratios of the integrated *dN/dy* yields. If we neglect the decay contributions and consider only the primordial yield, the anti-particle to particle ratios are found to be controlled only by their respective fugacities. i.e.,

$$\frac{n_i}{\bar{n}_i} = \exp\frac{\mu_i - \bar{\mu}_i}{T} = \exp\left(\frac{2\mu_i}{T}\right) \quad (2)$$

Ratios of particles with the same mass, but different quark content, such as $\bar{p}/p$ and $K^-/K^+$

are sensitive to the balance between matter and antimatter, characterized by the baryon chemical potential $\mu_B$. As strange quarks are created during the collision and are not transported from the incoming nuclei, strangeness production is expected to be a good estimator of the degree of equilibration of the produced fireball [8]. The $\bar{p}/p$ ratio in accordance with equation 2 can be written as:

$$\frac{\bar{p}}{p} = \exp(-2\frac{\mu_B}{T}) \tag{3}$$

The other particle ratios of thermal yields (i.e. without feed down contributions from the heavier resonances) can be correlated accordingly with the $\bar{p}/p$ ratio as:

$$\frac{K^-}{K^+} = \exp\left(2\frac{\mu_s}{T}\right)\left(\frac{\bar{p}}{p}\right)^{\frac{1}{3}} \tag{4}$$

$$\frac{\bar{\Lambda}}{\Lambda} = \exp(-2\frac{\mu_s}{T})\left(\frac{\bar{p}}{p}\right)^{\frac{2}{3}} \tag{5}$$

$$\frac{\bar{\Sigma}}{\Sigma} = \exp(-2\frac{\mu_s}{T})\left(\frac{\bar{p}}{p}\right)^{\frac{2}{3}} \tag{6}$$

$$\frac{\bar{\Xi}}{\Xi} = \exp(-4\frac{\mu_s}{T})\left(\frac{\bar{p}}{p}\right)^{\frac{1}{3}} \tag{7}$$

$$\frac{\bar{\Omega}}{\Omega} = \exp(-6\frac{\mu_s}{T}) \tag{8}$$

Incidentally, the above given relations of the mid-rapidity equal mass particle ratios, emitted from a hadronic fireball maintaining a high degree of thermal and chemical equilibration, hold even when the resonance decay contributions are included [8].

In our model [1, 7], it is assumed that the rapidity axis is populated with hot hadronic regions moving along the beam axis with monotonically increasing rapidity $y_0$. This essentially emerges from the situation where the colliding nuclei exhibit transparency effects. Hence the regions away from the mid-region also consist of the constituent partons of the colliding nucleons, which suffer less rapidity loss due to partial nuclear transparency. Due to this, these regions have an excess of quarks over the anti-quarks and hence maintain larger baryon chemical potentials on either side of the mid-region in a symmetric manner. For this reason, a quadratic-type rapidity dependent chemical potential $\mu_B$ has been considered in our model as,

$$\mu_B = a + by_0^2 \tag{9}$$

where the model parameter $a$ defines the chemical potential at mid-rapidity and the parameter $b$ gives the variation of the chemical potential along the rapidity axis. We focus on the mid-rapidity data (dN/dy), for which a bulk of published hadrons yields is available. We have also employed the strangeness conservation criteria in a way such that the total strangeness in the fireball is zero. We have tabulated below the different values of chemical potentials obtained in our previous papers [1, 7] for different centre of mass energies, as shown in table 1, by using our Unified Statistical Thermal freeze-out Model. For the sake of comparison, we have also mentioned the values of $\mu_B$ obtained at different SPS and RHIC energies by STAR collaboration [11] and by ALICE at LHC [12].

| $\sqrt{s_{NN}}$ (GeV) | $\mu_B = a$ (MeV) (Our Model) | $\mu_B$(MeV) Ref. [11,12] |
|---|---|---|
| 9.2 GeV | 290 ± 3 | 300 ± 12 |
| 62.4 GeV | 45 ± 5 | 62.7 ± 6 |
| 130 GeV | 25 ± 3 | 29 ± 4.6 |
| 200 GeV | 23 ± 2 | 22 ± 4.4 |
| 2.76 TeV | 0.5 ± 0.5 | 0.75 |

**Table 1**. The values of chemical potential obtained at different collision energies in our model are compared to the values obtained in References [11, 12].

In order to reproduce the variation of various particle ratios, at all possible energies up to the LHC, we need to obtain the dependence of the chemical potential and the chemical freeze-out temperature on the collision energies. For this purpose, we use the following parameterization [9]:

$$\mu_B = \frac{c}{1+d\sqrt{s_{NN}}} \quad (10)$$

Using the set of extracted values of $\mu_B$ from our model at five different energies we obtain $c$ = 1304 MeV and $d$ = 0.38 GeV$^{-1}$ when $\mu_B$ is in the units of MeV. These values are found to be in a close vicinity with the values of the parameters $c$ = 1308 ± 0.028 MeV and $d$ = 0.273 ± 0.008 GeV$^{-1}$ obtained by Cleymans et. al. [13, 14]. Similarly, in order to obtain the values of freeze-out temperatures at various energies, we first fit the $\bar{\mathrm{p}}/\mathrm{p}$ and $\bar{\Lambda}/\Lambda$ ratios at five different collision energies by using the corresponding values of the chemical potentials obtained through our model as mentioned in Table 1. The five different values of freeze-out temperatures obtained in this way are shown in Table 2. To obtain the chemical freeze-out temperature at all possible energies we use the following parameterization:

$$T = T_0\left[1 - \frac{1}{(\log\sqrt{s_{NN}} - e)/f}\right] \quad (11)$$

Using the set of chemical freeze-out temperature obtained at five different collision energies we obtain the values of the parameters in equation (11) as $T_0$ = 172 MeV, $e$ = 1.10 and $f$ = 0.14. The result of these parameterizations (eq. 10, 11) is shown below in figure 1. The solid red curves represent the best fit and the black square shapes represent the values of chemical potentials and the freeze-out temperatures as obtained previously in our model.

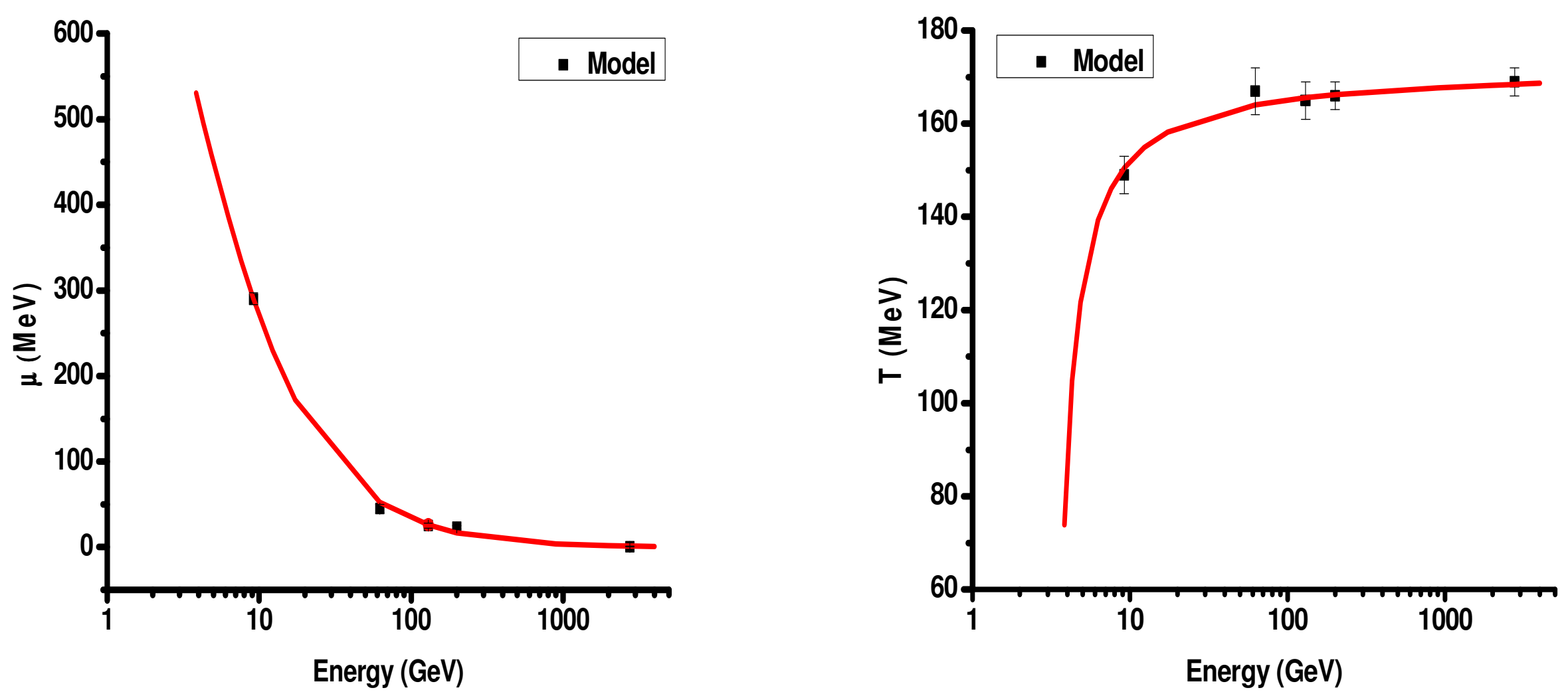

**Figure 1:** The energy dependence of baryon chemical potential (left) and chemical freeze-out temperature (right) in our model.

In the Table 2, we have also shown the values of kinetic freeze-out temperatures $T_{kin}$ obtained by reproducing the transverse momentum distributions of protons and kaons at the five different collision energies in our model [1, 7]. It is seen that the values of chemical freeze-out temperature T and the kinetic freeze-out temperature $T_{kin}$ are almost same at $\sqrt{s_{NN}}$ = 9.2 GeV while as at RHIC, the values of chemical freeze-out temperature T are a little higher than the values of kinetic freeze-out temperature $T_{kin}$ . At LHC, this difference is even larger. Thus it seems that the time duration between the two types of freeze-outs is dependent on the collision energy. This duration is larger for the larger collision energies. This may be understood as a result of the larger particle production (and hence a larger system size) at higher energies, which results in the development of the larger collective flow effects at the cost of thermal temperature. Thus the thermal freeze-out temperature is reduced considerably at LHC. On the other hand, at lower energies, the lesser particle production (and hence a smaller system size) does not allow the significant development in collective flow effects and hence the temperature drop is not significant. This may result in the almost similar values of the two freeze-out temperatures at lower energies.We find in figure 1 that the extracted temperature values generally increases rapidly where as the baryon chemical potential decreases monotonically with the collision energy and both tend to saturate at RHIC and LHC energies. In general, these values are found to lie close to the ideal gas values. Using these values of $\mu_B$ and T, we reproduce the energy dependence of various anti-particle to particle ratios by using the equations 3 – 8.

| $\sqrt{s_{NN}}$ (GeV) | $T$ (MeV) (Our Model) | $T_{kin}$ (MeV) (Our Model) |
|---|---|---|
| 9.2 GeV | 149 ± 4 | 150 ± 2 |
| 62.4 GeV | 167 ± 5 | 170 ± 5 |
| 130 GeV | 165 ± 4 | 163 ± 2 |
| 200 GeV | 166 ± 3 | 162 ± 2 |
| 2.76 TeV | 169 ± 3 | 103 ± 1 |

**Table 2.** The values of chemical and kinetic freeze-out temperatures obtained at different collision energies in our model.

These particle ratios are plotted in Figures 2 – 4 below. The experimental data shown by red colored shapes in Figures 2 and 3 is taken from [9] and the references there in, while for Figure 4, the data is taken from [10] and the references there in. In heavy-ion collisions the increase in the antimatter to matter ratio with the center of mass energy of the system has been observed earlier by the NA49 and the STAR collaborations. The increase of $\bar{p}/p$ ratio towards unity with an increase in the centre of mass energy from SPS to LHC is shown in the left panel of Figure 2. Our model results are in good agreement with the experimental data. The increase in $\bar{p}/p$ ratio with an increase in collision energy reflects the decrease in net baryon density towards higher collision energies thus making the collision system partially transparent at RHIC and almost completely transparent at LHC. The ratio $K^-/K^+$, plotted in right panel of Figure 2, shows a significant dependence on the centre of mass energy. Our model prediction shows a close agreement with the experimental data points. Hence the overall feature of the experimental data in both cases is in good agreement with our model calculations. The decrease in $K^-/K^+$ ratio with a decrease in collision energy is due to an increase in net baryon density which leads to the associated production of kaons thus favoring the production of $K^+$ over $K^-$. In Figure 3 and in the left panel of Figure 4, we have plotted the energy dependence of singly, doubly and triply strange anti-baryon to baryon ratios. We find a fairly good agreement between our model results and the experimental data. It is seen that the ratios increase towards unity with the increase in collision energies. The ratios appear ordered with the strangeness quantum number, i.e. the higher the strangeness quantum number, the smaller the difference between anti-baryon and baryon content. This is so called "mass-hierarchy" where the saturation value (i.e.1) of the ratio is achieved earlier for the more massive hyperon species [10]. It is interesting to note that even the yield of rarely produced strange particles like Ω is also fairly well described by our USTF model. In the right panel of Figure 4, we have shown the energy dependence of $\pi^-/\pi^+$ ratio. This is also reproduced with a fairly good agreement. It is seen that anti-baryon/baryon ratios have more sharp dependence on collision energy.

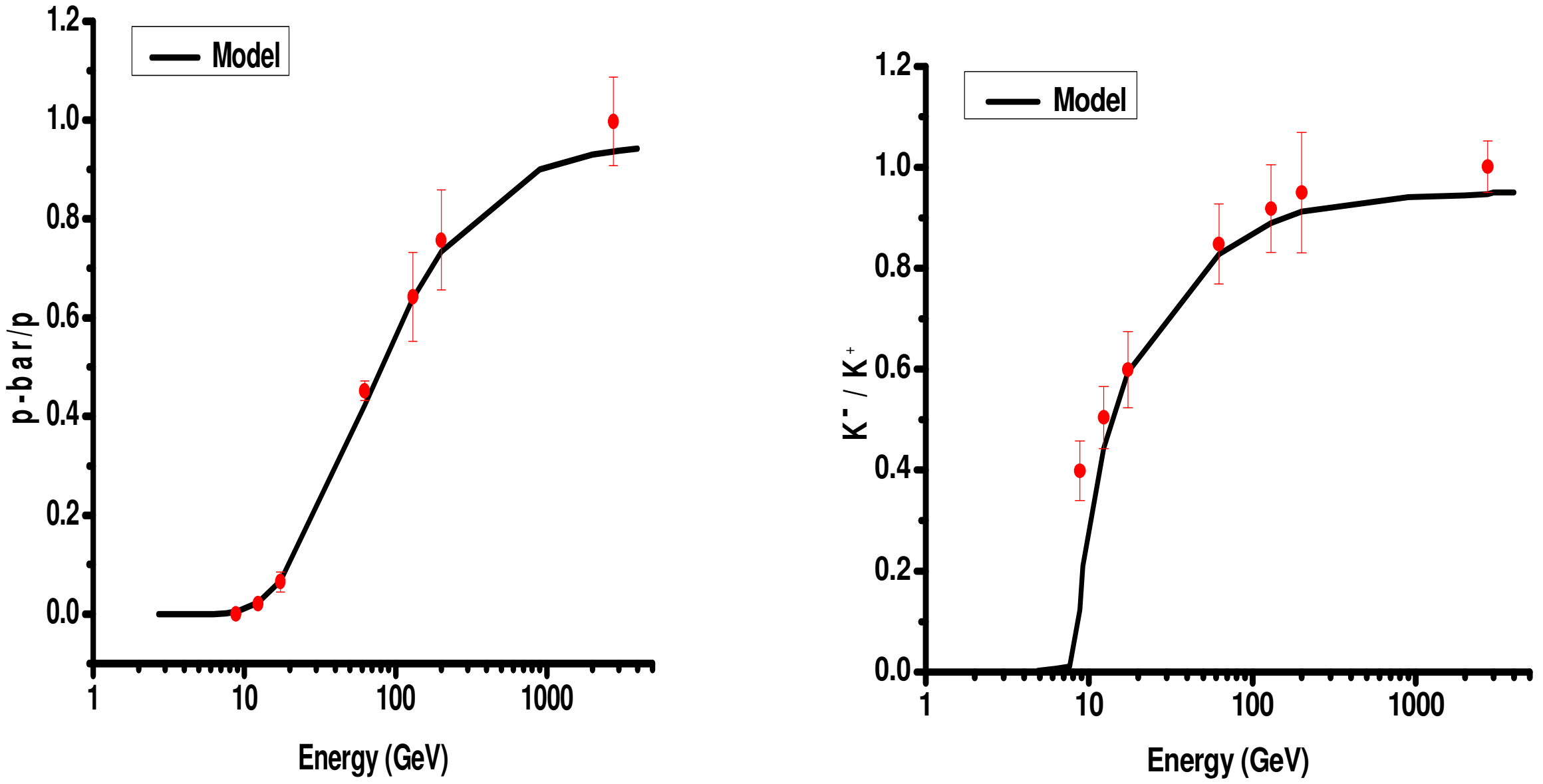

**Figure 2**: Energy dependence of $\bar{p}/p$ (left) and $K^-/K^+$ (right) ratios.

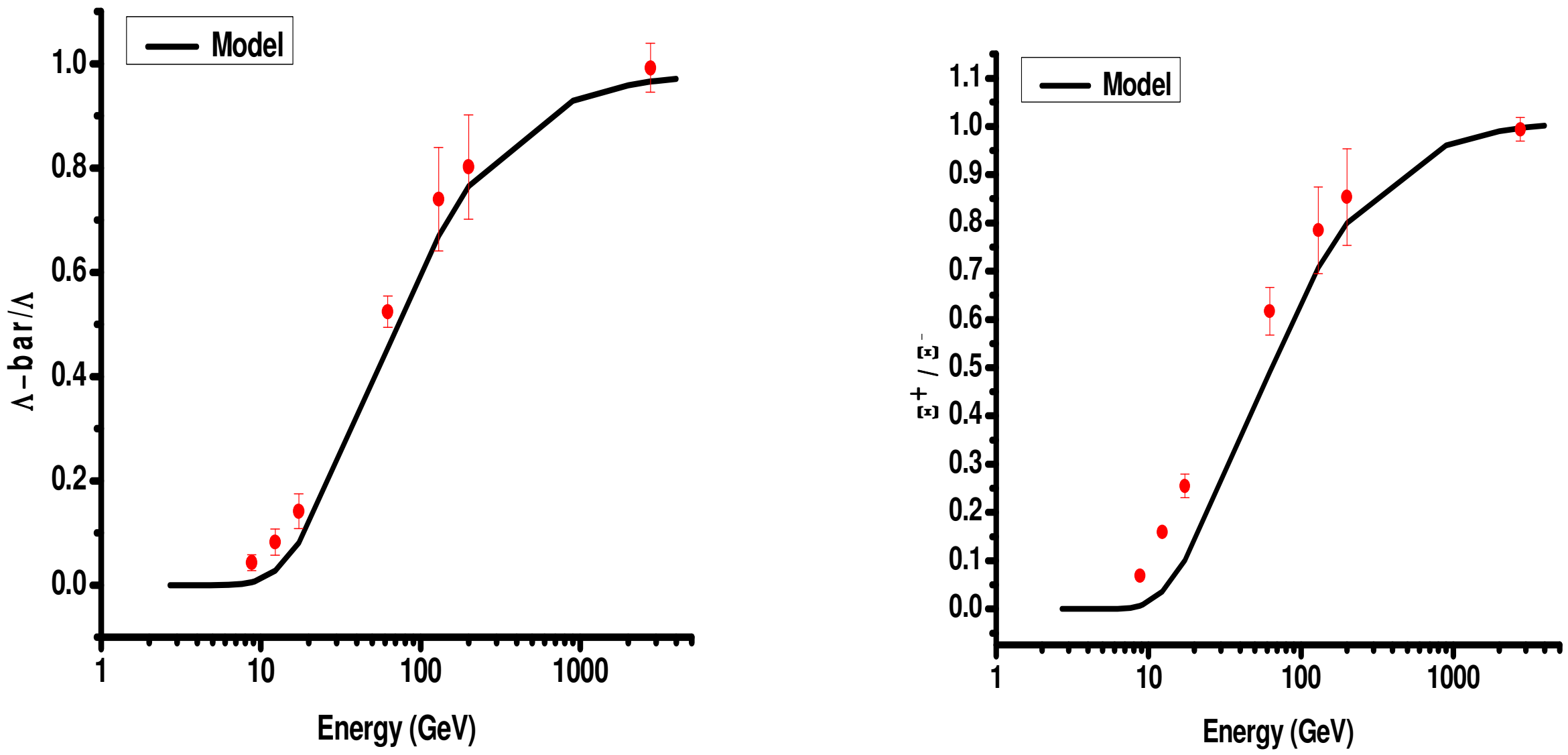

**Figure 3:** Energy dependence of $\frac{\bar{\Lambda}}{\Lambda}$ (left) and $\frac{\bar{\Xi}}{\Xi}$ (right) ratios.

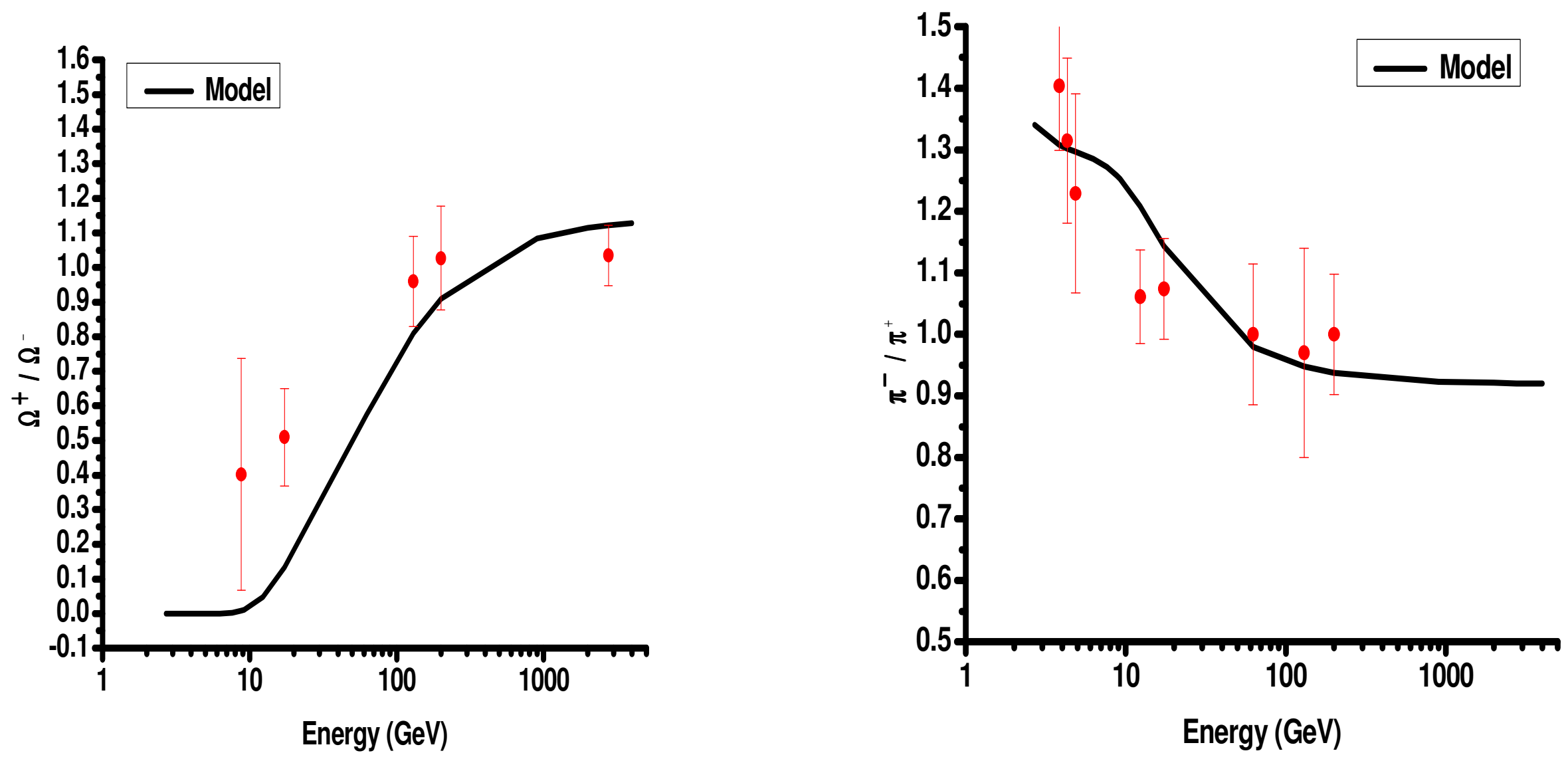

**Figure 4:** Energy dependence of $\frac{\bar{\Omega}}{\Omega}$ (left) and $\frac{\pi^-}{\pi^+}$ (right) ratios.

Figure 5 shows the correlation of the $K^-/K^+$ (representing net strange chemical potential $\mu_s$ ) with the $\bar{p}/p$ ratio (representing net baryon chemical potential $\mu_B$). The black solid curve, which is a result of the equation 4 ( $\frac{K^-}{K^+} \sim (\frac{\bar{p}}{p})^{\frac{1}{3}}$ ) represents our model prediction where as the colored shapes represents the different experimental data points as mentioned in the Figure. This correlation could give information on how the kaon production is related to the net-baryon density. At lower energies, the kaon production is dominated by the associated production which results in more $K^+$ production compared to $K^-$. Also the $\bar{p}/p$ ratio is much less than unity, indicating that there is a large baryon stopping at the lower energies. As we go towards higher energies, the pair production mechanism starts to dominate and the ratios tend to become closer to unity.

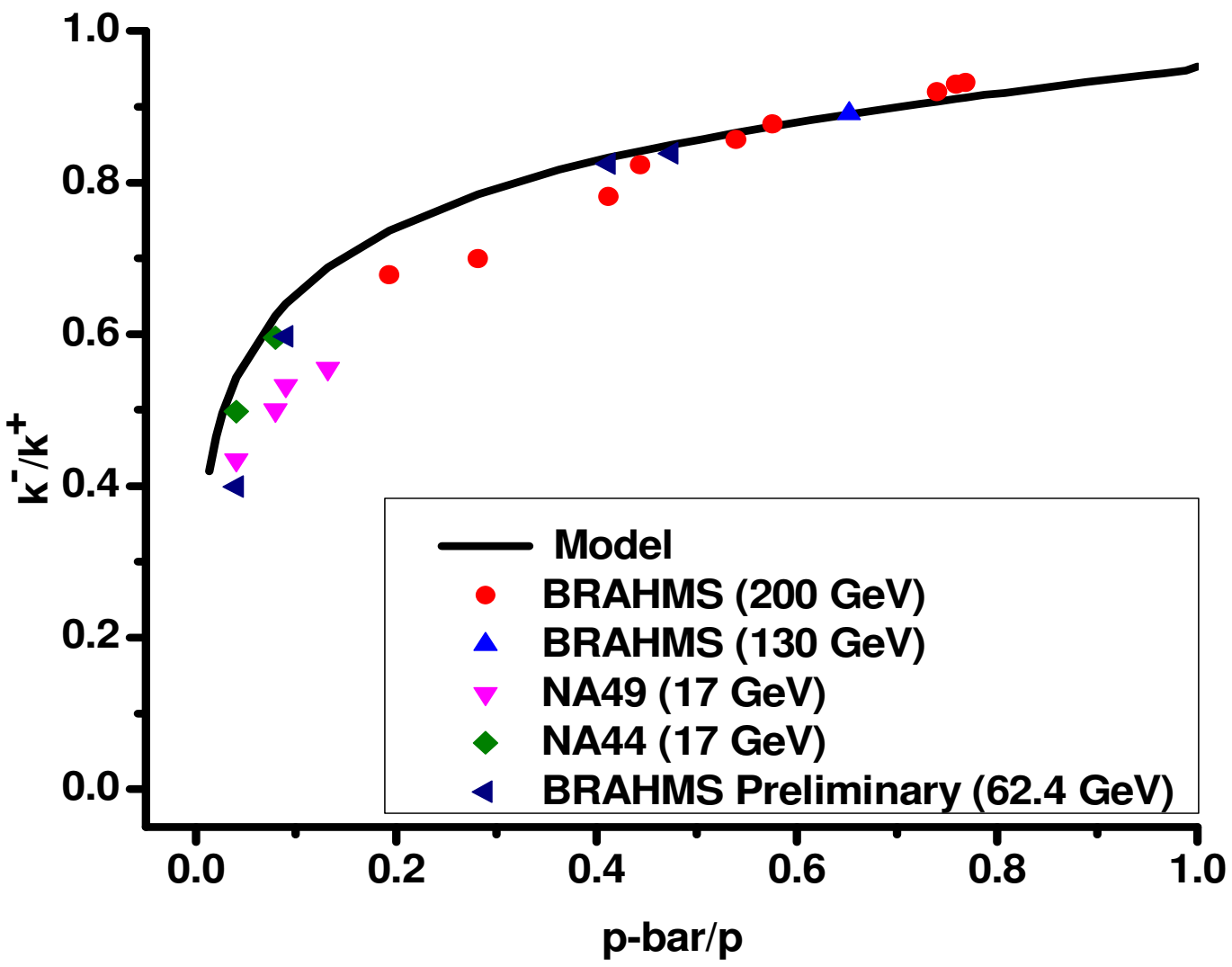

**Fig.5** Correlation of $K^-/K^+$ ratio with $\bar{p}/p$ ratio at mid-rapidity for central collisions.

We have further reproduced the energy dependence of various other particle yields relative to pions as shown in Figure 6. The experimental data is taken from [9, 10] and the references there in. Unlike in case of anti-particle to particle ratios, here we have to include the effects of resonance decay contributions, as discussed in the first section. Our model results are seen to be in a fair agreement with the experimental data points. At lower energies, a peak has been observed in $K^+/\pi^+$ ratio at around 8 GeV [13, 15, 16]. This is not reproduced well in our analysis because of the invalidity of our approach at these lower energies. The $K^-/\pi^-$ ratio exhibits no sharp structure and instead a smooth evolution with the collision energy is seen. Our model results for the $K^-/\pi^-$ ratio slightly over predict the experimental data points at lower energies. However, the main features of the data showing a steady decrease towards lower energies is well reproduced. Also we find that all the particle ratios seem to saturate at RHIC and LHC energies. This saturation seems to arise due to almost constant chemical freeze-out temperatures at RHIC and LHC energies. The steep decrease of the p/$\pi^+$ ratio towards higher energies reflects a decrease in the baryon chemical potential and hence an increase in the nuclear transparency, though the increase of pion production also plays a role in this. Beyond $\sqrt{s_{NN}}$ = 100 GeV the flattening of the curves takes place.

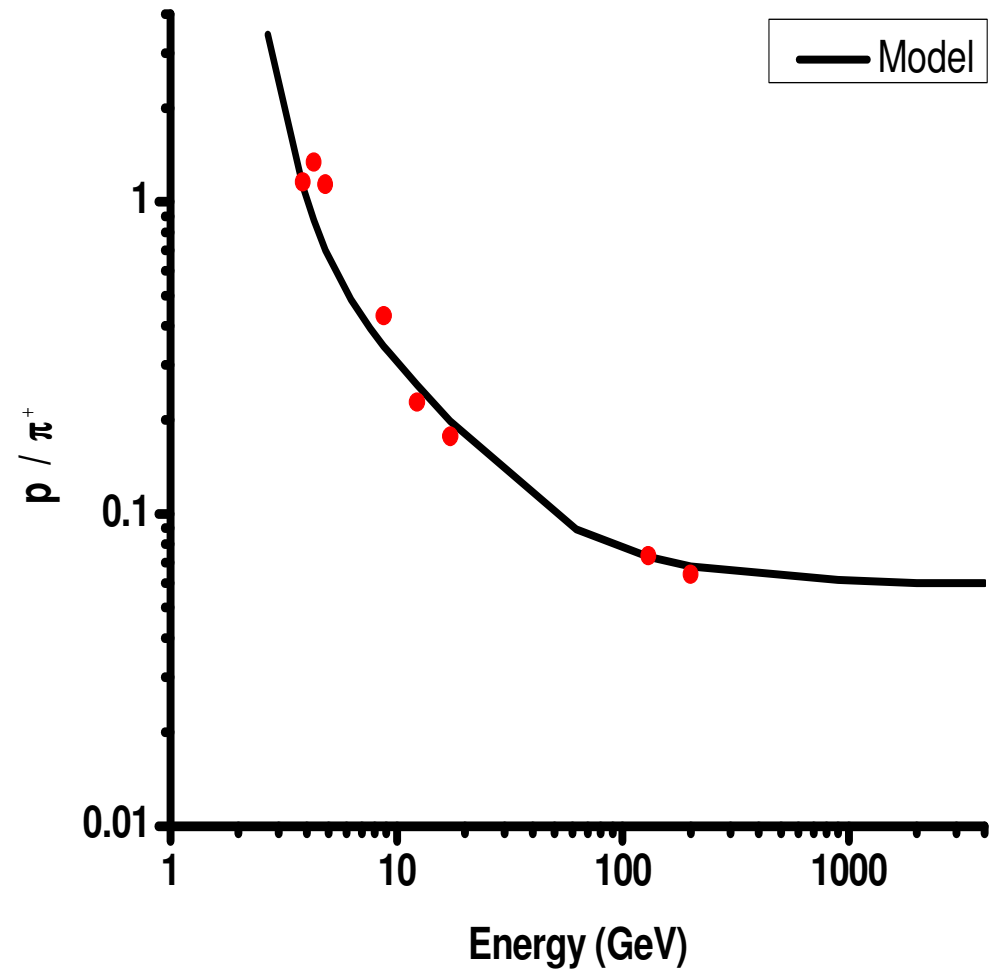

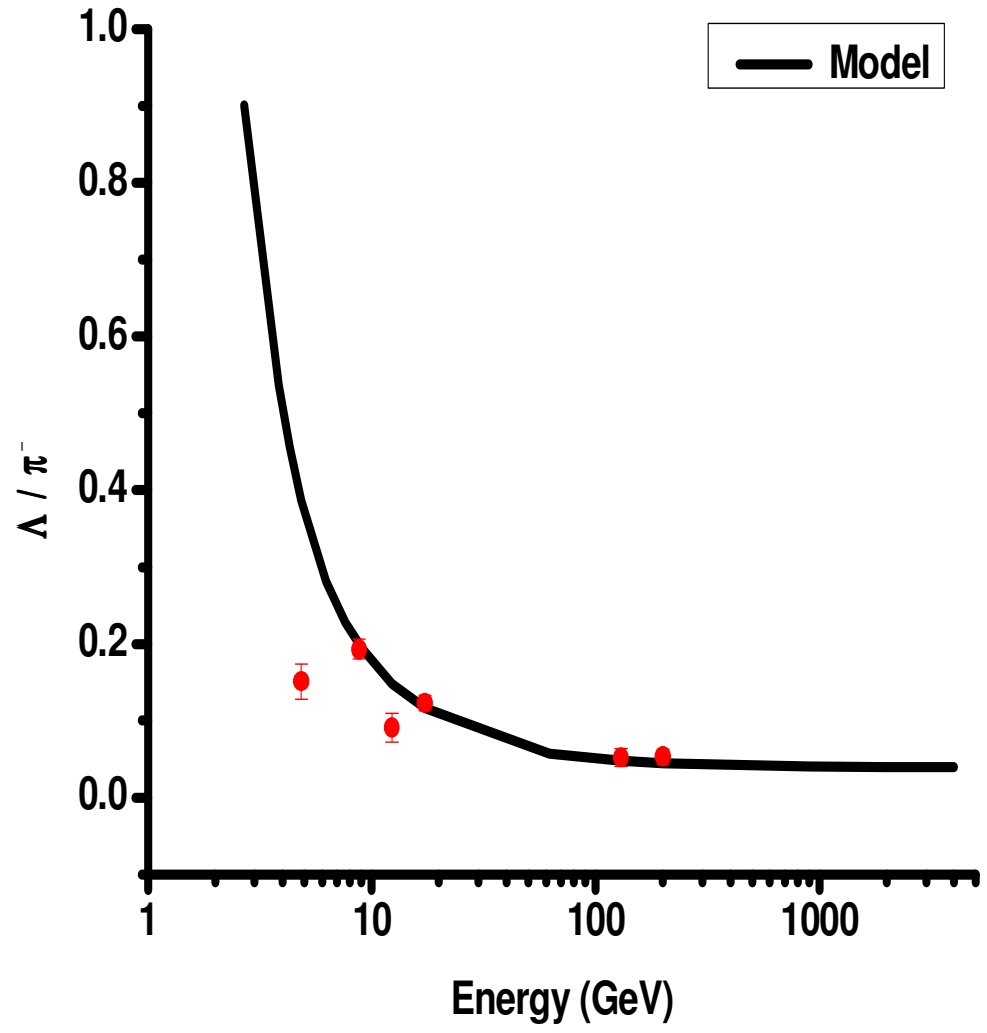

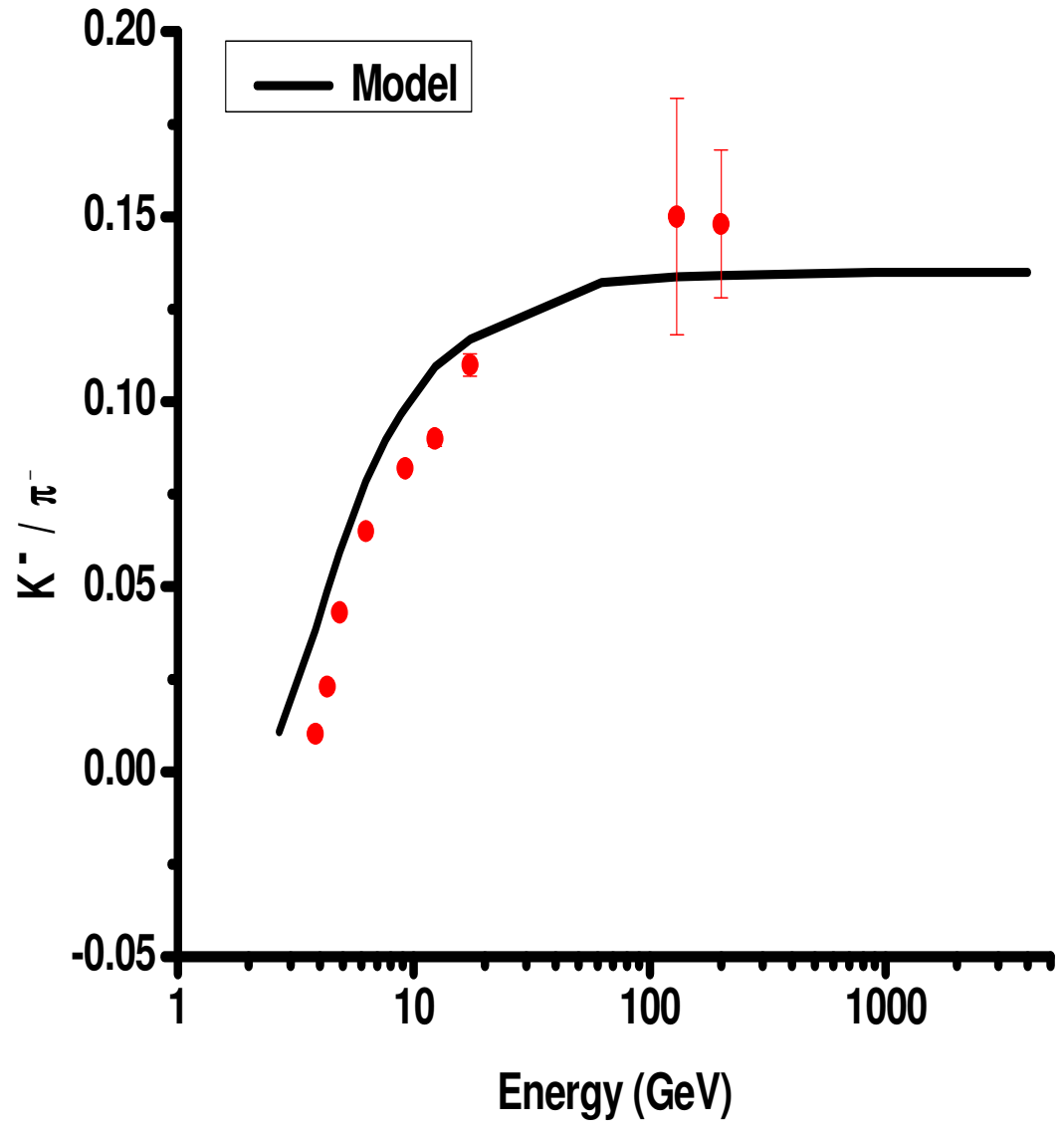

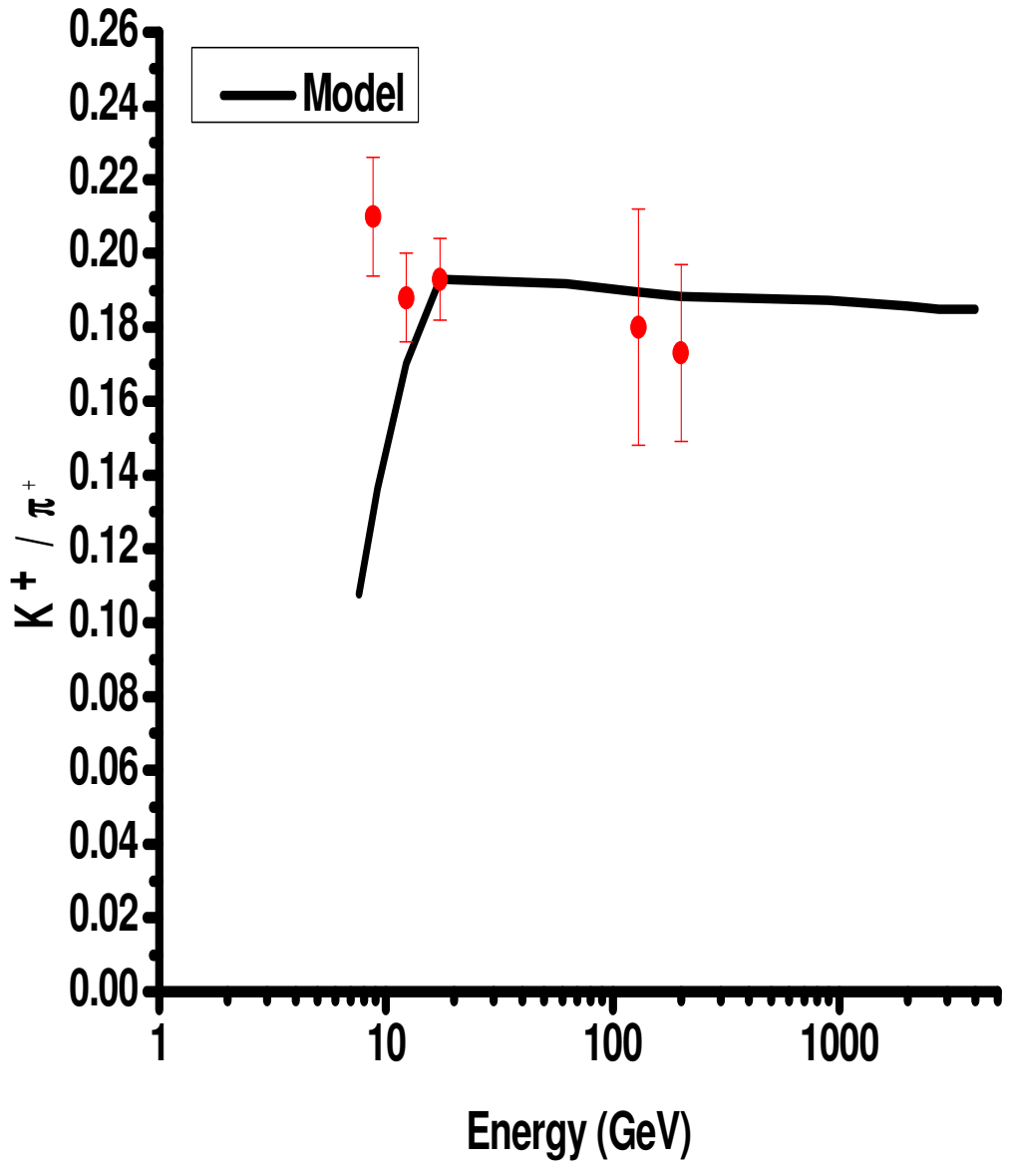

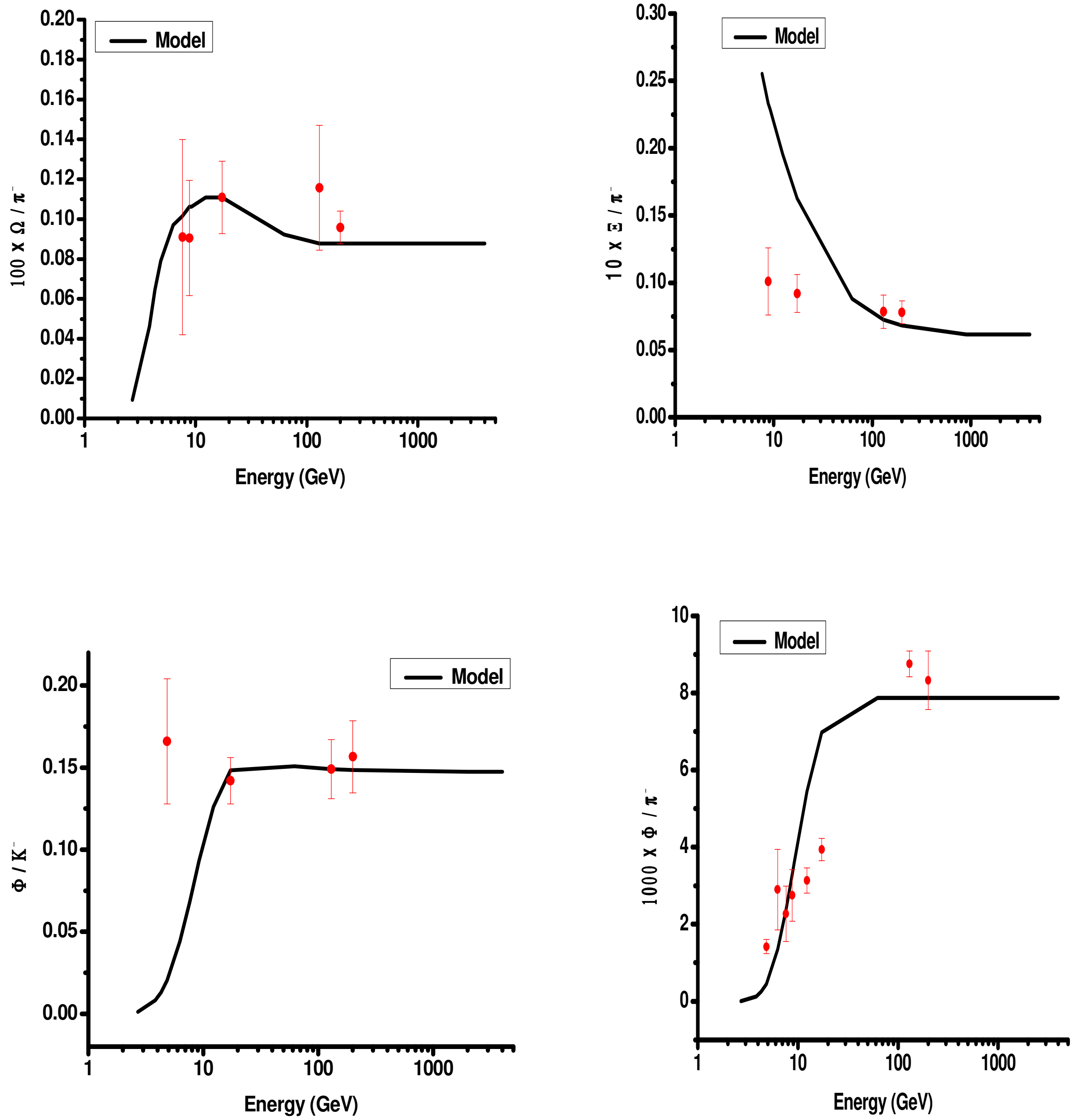

**Figure 6**. Energy dependence of various unlike particle ratios.

We have predicted the various hadron ratios produced in central Pb-Pb collisions at LHC and are presented in Table 3 below. The anti-particle by particle ratios are close to unity reflecting a very small chemical potential at LHC. Also the yields relative to pions are very similar to the values measured at RHIC energy.

| $K^-/K^+$ | $\bar{p}/p$ | $\bar{\Lambda}/\Lambda$ | $\Xi^+/\Xi^-$ | $\bar{\Omega}/\Omega$ | $\pi^-/\pi^+$ | $p/\pi^+$ | $K^+/\pi^+$ | $K^-/\pi^-$ | $\Lambda/\pi^-$ |
|---|---|---|---|---|---|---|---|---|---|
| 0.947 | 0.936 | 0.966 | 0.996 | 1.121 | 0.94 | 0.060 | 0.184 | 0.135 | 0.039 |
| $\Xi^-/\pi^-$ | $\Omega^-/\pi^-$ | $\varphi/K^-$ | | | | | | | |
| 0.061 | 0.087 | 0.147 | | | | | | | |

**Table. 3:** Model predictions for hadron ratios at LHC (T=168.65 MeV, $\mu_B$ = 0.85 MeV). Decay contributions are included.

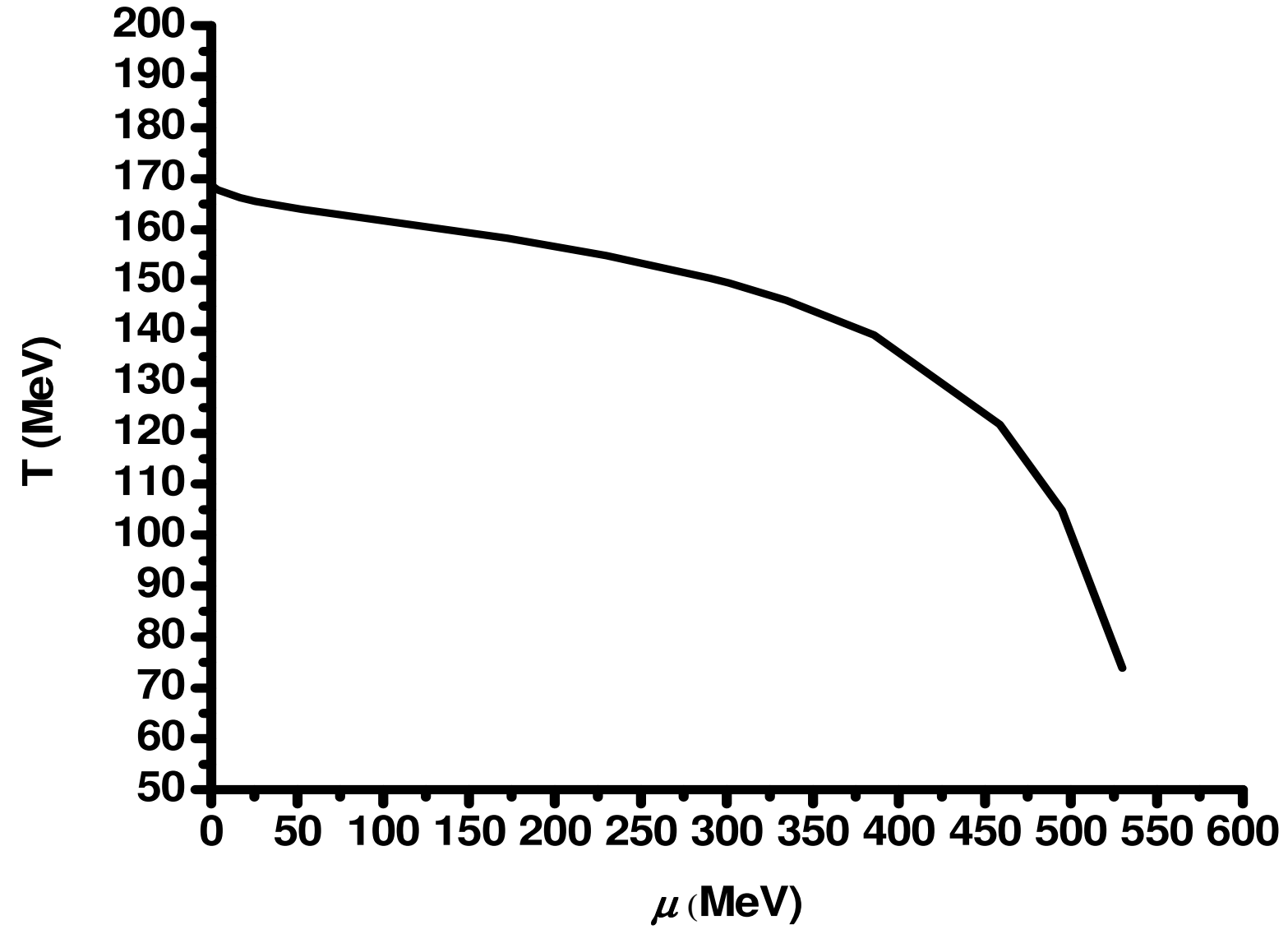

**Figure 7:** Chemical freeze-out temperature vs baryon chemical potential at different collision energies.

In Figure 7, we have plotted the chemical freeze-out temperature on the vertical axis and the chemical potential on the horizontal axis obtained at different collision energies. The value of chemical freeze-out temperature at RHIC and LHC lies in the vicinity of lattice QCD predicted phase transition temperature of ~ 170 MeV [9], indicating that the freeze-out occurs almost simultaneously after the phase transition.

## Summary and Conclusion

We have used our model (USTFM) to analyze the variations of the ratios of particles, produced in the high energy nucleus-nucleus collisions, with center-of-mass energy and their correlations. We have compared our results with the experimental data. A good agreement between our model results and experimental data shows that thermal model used in the analysis gives a satisfactory description of the data and hence the validity of our approach. The dependence of the baryon chemical potential and the chemical freeze-out temperature on the energy is studied. For this purpose a parameterization is used for each case. It is found that the extracted value of chemical freeze-out temperatures at RHIC and LHC are almost constant and are close to the lattice QCD predicted phase-transition temperature, suggesting that chemical freeze-out happens in the vicinity of the phase boundary i.e. shortly after hadronization process is completed. Also the difference between the chemical freeze-out and the kinetic (thermal) freeze-out temperatures is found to increase with the collision energy.

## Acknowledgements

We acknowledge the financial support from University Grants Commission (UGC) for this work.